\documentstyle[prd,aps,preprint]{revtex}
\begin{document}
\draft

%
%

\preprint{Nisho-00/2} \title{Ultra High Energy Cosmic Rays 
and Gamma Ray Bursts\\
from Axion Stars } 
\author{Aiichi Iwazaki}
\address{Department of Physics, Nishogakusha University, Chiba
  277-8585,\ Japan.} \date{August 30, 2000} \maketitle
\begin{abstract}
We propose a model in which ultra high energy cosmic rays 
and gamma ray bursts are produced by 
collisions between neutron stars and axion stars.
The acceleration of such a cosmic ray is made by the electric field, 
$\sim 10^{15}\,\,(B/10^{12}\,\mbox{G})\,\,\mbox{eV}\,\,\mbox{cm}^{-1}$, 
which is induced in 
an axion star by relatively strong magnetic field $B>10^{12}$ G 
of a neutron star. 
On the other hand,
similar collisions generate gamma ray bursts when magnetic field
is relatively small, e.g. $\leq 10^{10}$ G. 
Assuming that
the axion mass is $\sim 10^{-9}$ eV, 
we can explain huge energies 
of the gamma ray bursts $\sim 10^{54}$ erg as well as 
the ultra high energies of the cosmic rays $\sim 10^{20}$ eV.
We estimate rate of energy release in the collisions and we find that
the rate roughly agrees with observations.
In addition, we show that
these axion stars are plausible candidates for MACHOs.
Since the axion star induces oscillating electric current under the 
magnetic field, 
observable monochromatic radiations are
emitted.

\end{abstract}

\vskip .7cm

\pacs{14.80.Mz, 95.30.+d, 98.80.Cq, 98.70.Rz, 97.60.Jd, 05.30.Jp, 98.70.Sa, 
11.27.+d
\hspace*{3cm}}
\vskip2pc
\tightenlines
\section{Introduction}
The origin of ultra high energy cosmic rays ( UHECRs ) is 
one of most mysterious puzzles in astrophysics\cite{uhe}.
UHECRs with energies $\sim 10^{20}$ eV can not travel in distance 
more than $50$ Mpc 
owing to the interactions between UHECRs 
and the cosmic background radiations\cite{gzk}.
Observations, however, show that there are no visible candidates for 
the generators of such UHECRs in the arrival directions of UHECRs.  
There are several models of generators for UHECRs, but none of them
give satisfactory explanations. 
They are divided into two categories\cite{2006}.
One is conventionally astrophysical models, and the other one
Are particle physical models.  
It is explained in astrophysical models
that the generator could be an object such as neutron star,
active galactic nuclei or the same astrophysical object 
as generators of gamma ray bursts.
It is explained in particle physical models that UHECRs could be caused by
hypothetical objects such as cosmic string\cite{costring}, 
magnetic monopole\cite{monopole}, 
super heavy relic particles decaying today\cite{relic}, e.t.c.. 
The astrophysical models
have a difficulty that cosmic rays can not be accelerated to 
such high energies.
On the other hand, the particle physical model has difficulties
concerning to production rate of UHECRs and a large number 
of theoretical assumptions
on hypothetical objects.

The origin of gamma ray bursts ( GRBs) is also one of most 
mysterious puzzles in astrophysics\cite{grb}. Some of their energies reach 
$\sim 10^{54}$ erg in the case of spherical emission of the burst. 
It is very difficult to 
find a model explaining such a huge energy. 
There are, however, several
models proposed. They are also divided into two categories, astrophysical
one and particle physical one. Since conventional astrophysical 
explanation are optimistically believed to have no serious problems,
particle physical one is not urgently needed so that no such serious model
is pursued. Conventionally, astrophysicists consider 
that the generator could be
merger of neutron star-neutron star, neutron star-black hole or the 
collapstar ( collapse of extremely massive star ), e.t.c.G
recently, there are a number of growing evidences supporting that
GRBs with long duration are caused by the collapse of extremely
massive star.  
These models
are optimistically believed to explain the properties of GRBs
such as amount of energies released, 
baryon contamination of eject, afterglow spectrum, e t c.. But owing to 
difficulties in the treatment of MHD involving neutrinos and 
unknown properties in the fluid of nucleons, 
there is no successful model among them. In particular, 
the problem of the baryon contamination seems difficult to be solved
in conventional astrophysical models. 
On the other hand, in some particle physical models, axions or 
other unknown weakly interacting particles are used 
to overcome the baryon contamination problem 
based on the conventional astrophysical engines.  
Moreover, in other 
particle physical models, one claims that 
the engine is collapse of mirror star\cite{mirror}, cosmic string\cite{stringrb}
or collision between axion star and neutron star\cite{iwazaki}.  
The engine in these model is associated with
dark matter; mirror matter, cosmic string or axion.

Among these candidates for dark matter, the axion\cite{PQ,text,kim} is 
the most plausible one.  
Probably, some of axions may form boson stars ( axion stars ) 
in the present Universe
by gravitational cooling\cite{cooling} or 
gravitational collapse of axion clumps formed at the period of 
QCD phase transition\cite{kolb}. In this paper 
we explain a generation mechanism\cite{iwazaki}
of UHECRs and GRBs, and discuss their production rates. 
Our mechanism for both phenomena is collision between axion stars 
and neutron star. Different strength of the magnetic field 
at the surface of the neutron star yields these different 
phenomena, i.e. UHECR and GRB.
Furthermore, the axion stars can be plausible candidates 
for gravitational microlenses ( MACHO\cite{MACHO} ) with an appropriate
choice of the axion mass; the choice is taken in our model for explaining
huge energies $\sim 10^{54}$ erg in GRBs and 
extremely high energies $\sim 10^{20}$ eV in UHECRs. 
It is intriguing that the masses of the axion stars 
become those of the candidates for MACHOs
with this choice of the axion mass. 
We also show that the collisions 
cause emission of observable monochromatic radiations from the axion stars\cite{iwaz}. 
We can check the validity of our model by detecting the radiations.

We have previously proposed\cite{iwazaki} a possible generation mechanism 
of gamma ray burst ( GRB ).
According to the mechanism 
the collision between  
axion star and neutron star 
generates a gamma ray burst; the axion star 
dissipates
its mass energy\cite{iwazaki,iwa} very rapidly under the strong magnetic field 
of the neutron star. Thus, the energy released in the collision is given by 
the mass $M_a$
of the axion star. Typically, 
$M_a\sim 10^{-5}M_{\odot}\,(10^{-5}\mbox{eV}/m_a)$ where
$m_a$ ( $M_{\odot}$ ) denotes the axion mass ( solar mass ). 
In the previous analysis we have taken 
the mass, $m_a\sim 10^{-5}$ eV, as suggested observationally
in standard invisible axion models.  The mass of axion star, 
$M_a\sim 10^{49}$ erg, corresponding to this choice, however,   
is not enough to explain a huge energy $\sim 10^{54}$ erg 
observed in GRB980123 even if moderate jet is assumed in the GRB:
If the solid angle of the jet is
given by $\Omega$, the energy released in the GRB is about 
$10^{54}\,\Omega/4\pi$ erg.

In the present paper, we  
assume that the axion mass is given by  $m_a\simeq 10^{-9}$ eV,
although the choice is not conventional\cite{text,string}. 
Then, we can explain that 
the energies of GRBs can reach  
the huge energies $\sim 10^{54}$ erg. We can also show\cite{iwazaki} that  
cosmic rays are accelerated to the energies $\sim 10^{20}$ eV.

The essence of both phenomena is that  
a strong electric field 
$\sim 10^{15}\,\,(B/10^{12}\mbox{G})\,\,\mbox{eV}\,\,\mbox{cm}^{-1}$
is induced in the axion star when it is located under magnetic field $B$
of a neutron star. This electric field can accelerate
charged particles to the huge energies $\sim 10^{20}$ eV
in the case of $B>10^{12}$ G, although such a strong electric field
is unstable against electron positron pair productions.
On the other hand, weaker and more stable electric field dissipates 
its energy in magnetized 
conducting medium of neutron star. In particular, in the case of neutron 
stars with relatively weak magnetic field $\leq 10^{10}$ G, the 
axion star collides directly with them and its whole energy
is dissipated very rapidly in their outer crust owing to the energy dissipation
of the electric field. This rapid energy dissipation
produces jet of baryons, which leads to emission of GBRs.
In the case of strong magnetic field $>10^{12}$ G, however, the axion star
evaporates before colliding directly with the neutron star; the evaporation
occurs owing to the instability of such a strong electric field 
against electron-positron pairs creation\cite{vacdecay}. 
Therefore, the axion star decays into
electron positron pairs which are accelerated to the energies 
$\sim 10^{20}$ eV by the electric field during its decay. 
In both cases the energies released are given by the mass of the axion
star. 
Since the mass of the axion star is given by  
$\sim 10^{-1}M_{\odot}\simeq 10^{53}$ erg with 
the assumption of $m_a\simeq 10^{-9}$ eV,     
the energy $\sim 10^{54}$ erg observed in some GRBs
can be explained by assuming a moderate jet of the GRBs.
Thus the collisions 
between the axion stars and the neutron stars are possible sources of 
both UHECRs and GRBs. 
Additionally, it turns out that the axion 
star is a plausible candidate for MACHO\cite{MACHO} 
because the value of the mass $M_a$  
is suitable for explaining the
observations of MACHOs. Since all of baryonic
candidates for MACHOs seem to have serious difficulties\cite{non},
nonbaryonic ones like the axion stars would be favored.

We also discuss that 
the collisions generate monochromatic radiations
with a frequency $m_a/2\pi\simeq 2.4\times 10^5$ Hz. Their flux 
is sufficiently large to be observed.
With the detection of
such radiations we can test our model and determine the axion mass.

It is reasonable to think that some of axion stars 
gravitationally attract matters,
i.e. H and He in the Universe 
and have baryon contamination, while the other ones
have no such contamination. Then, the axion stars with the baryon 
contamination produce protons as UHECRs
in the collision with neuron stars. On the other hand 
the other ones produce photons as UHECRs, which are generated in
fireball of electron-positron pairs.
Therefore, UHECRs produced in our mechanism are 
either protons or photons.


In the section ( 2 ) we describe the properties of the axionic boson stars
for latter conveniences. In the section ( 3 ) we explain how the axion star
induces an electric field under an external magnetic field. In particular,
we show that the axion star is electrically polarized in spite of 
the neutrality of the axion itself. In the section ( 4 ) we describe 
how the electric field can accelerate charged particles to the energies
$\sim 10^{20}$ eV. Such a strong electric field can be 
produced only around a neutron star with magnetic field $>10^{12}$ G. 
In the section ( 5 ) we explain a generation mechanism
of GRBs by noting that the electric field dissipates rapidly 
its energy in conducting medium of neutron star. In the section ( 6 ) 
we show that our generation mechanism of UHECRs and GRBs 
gives the rate of the energy release consistent 
with the observations. 
In the section ( 7 ) we point out an observational signal of 
our mechanism, that is, electromagnetic radiations 
from axion star rotating around neutron star. The existence of such an 
observational signal 
is distinctive feature of our model.
We summarize our results in the final section ( 8 ).

\section{Axion Stars}
Let us first explain axion stars ( ASs ), which are 
ones of most realistic boson stars.
The axion is a Nambu-Goldstone boson associated with Pecci-Quinn
symmetry, which was introduced for solving strong CP problem. 
The mechanism is the most plausible one for the problem.
The symmetry is broken spontaneously at large energy scale, 
$f_{PQ}\sim 10^{12}$ GeV.
But since the symmetry is anomalous, it is also broken dynamically 
by the effects of QCD instantons. Therefore, the axion obtains 
a mass, $m_a$.

The AS is a coherent object of the real scalar field 
$a(x)$ describing the axion. It is a gravitational bound state of axions
represented by a solution\cite{real,iwasol}
of the equation of the axion field coupled with gravity.
An approximate form of the solution\cite{iwasol} without nodes is obtained
in the limit of small mass $M_a<<M_{\odot}$,

\begin{equation}
\label{a}
a(x)=f_{PQ}a_0\sin(m_at)\exp(-r/R_a)\,, 
\end{equation}
where $t$ ( $r$ ) is time ( radial ) coordinate and 
$f_{PQ}$ is the decay constant of the axion; 
there are solutions with nodes but their energies are much higher than 
that of this solution. 
The value of $f_{PQ}$ is constrained conventionally 
from cosmological 
and astrophysical considerations\cite{text,kim} such that 
$10^{10}$ GeV $< f_{PQ} <$ $10^{12}$ GeV. The axion mass $m_a$ is 
given in terms of $f_{PQ}$ such that $m_a\sim 10^{-2}\,\,\mbox{GeV}^2/f_{PQ}$.
But, when we assume entropy productions 
below the temperature $1$ GeV in the early Universe, we may be 
released from the constraints\cite{entropy}.
Hereafter we assume that $f_{PQ}\sim 10^{16}$ GeV or $m_a\sim 10^{-9}$ eV
for explaining huge characteristic energies of UHECRs and GRBs.

In the formula,
$R_a$ represents the radius of 
an AS which has been obtained\cite{iwasol} numerically 
in terms of mass $M_a$ of the AS;

\begin{equation}
\label{R}
R_a=6.4\,m_{pl}^2/m_a^2M_a\simeq 1.6\times 10^5\,\mbox{cm}\,
m_9^{-2}\,(10^{-1}M_{\odot}/M_a),
\end{equation}
with $m_9=m_a/10^{-9}$ eV and $m_{pl}$ is Planck mass.
Similarly, the amplitude $a_0$ in eq(\ref{a}) is represented such that  

\begin{equation}
a_0=1.73\times 10^2\,(10^{5}\,\mbox{cm}/R_a)^2\,m_9^{-1}\,\,\,.
\end{equation}

Therefore, 
we find that the solution is parameterized by one free parameter,
either one of the mass $M_a$ or the radius $R_a$: Once we choose 
either of mass $M_a$
or radius $R_a$ of an AS, the whole properties of the AS are determined only 
in terms of the axion mass $m_a$ 
( $M_a$ and $R_a^{-1}$ should be less than a critical 
value mentioned soon below ).

It is also important to note that the solution is not static but
oscillating with the frequency of $m_a/2\pi$. 
It has been demonstrated\cite{re} 
that there is no static 
regular solution of the real scalar massless field coupled with gravity.
This may be general
property of the real scalar massive field like the axion field. 
In fact, the oscillation in the solutions cause emission of electromagnetic
radiations under a magnetic field, which is a general phenomenon 
expected in axions exposed to the magnetic field. 
On the other hand, 
static solutions\cite{re} of boson stars
exist in the complex scalar field.

We should comment that the solution (\ref{a}) can represent approximately 
solutions even with larger masses.
More general solutions\cite{real} with larger masses have higher oscillation
modes such as $\sin(3m_at),\,\sin(5m_at)$ e t c.. But their
amplitudes are much smaller than $a_0$.

The AS mass is determined by physical conditions under which the AS
has been formed; how large amount of cloud of axions are cooled gravitationally
to form the AS, etc.. The situation is similar to other stars such as
neutron stars or white dwarfs. A typical mass scale in these
cases is the critical mass\cite{star}; 
stars with masses larger than the critical 
mass collapse gravitationally into more compact ones or black holes. 
In the case of the AS, there also exists a critical mass $M_c$ 
which is given by\cite{real}

\begin{equation}
M_c\simeq 10^{-1}M_{\odot}\,m_9^{-1}\,\,. 
\end{equation}
The ASs with masses larger than this one collapse into black holes.
Therefore, we may adopt this critical mass $M_c$ 
as a characteristic mass scale of the ASs. 
The corresponding radius $R_a$ of the ASs 
with this critical mass
is given such that $R_a\simeq 1.6\times 10^{5}\,m_9^{-1}\,\mbox{cm}$. 
( The critical mass is the maximal mass, which ASs can take. Thus, 
their masses, in general, are smaller than this one. 
Since energies of ASs released
in GRBs are given by their masses, the maximal energy  
in GRBs is given by the critical mass ).

We wish to comment that a critical mass for stars
like neutron stars, white dwarfs etc., represents order of 
masses they possess actually in the Universe. For example, 
the masses of the neutron
stars are in a region of $1.0 M_{\odot}\sim 1.6 M_{\odot}$, while
the critical mass of neutron star is given by about $2 M_{\odot}$.
Therefore, it is a reasonable assumption that the characteristic mass of
the axion star is the order of the critical mass, $ M_c$.

Although the gravitational cooling for ordinary
star formation is in general 
ineffective because it is too slow process, it has been shown\cite{cooling} 
that the cooling is very effective for the real scalar axion field.
Thus, the axion stars can be 
easily formed gravitationally in a gas of the axions.
It is reasonable to assume that the most of the axions in the Universe
forms the axion stars.

Until now, we do not consider a case that the axion star has an
ordinal matter, i.e. H and He. It is very natural to 
consider a possibility that the axion 
attracts gravitationally such matters after its formation. In the case 
the star is composed of the axions and the matters. The amount of
the matters would be no more than the critical mass of the axion,
unless it collapses into a black hole. It turns out below that
such axion stars with matter contamination play role for
emitting protons as UHECRs in the collision with neutron stars.

\section{Axion Star under external magnetic field}
Let us explain how an AS induces an electric field 
under an external magnetic field $\vec{B}$, in particular, 
that of a neutron star. Owing to the interaction 
between the axion and the electromagnetic field
described by

\begin{equation}
\label{int}
L_{a\gamma\gamma}=c\alpha a(x)\vec{E}\cdot\vec{B}/f_{PQ}\pi \,\,\,, 
\end{equation}
where the value of $c$ ( of the order of unity )
depends on axion models\cite{DFSZ,hadron,kim},
the Gauss law is modified\cite{Si} such that 

\begin{equation}
\vec{\partial}\vec{E}=-c\alpha \vec{\partial}\cdot(a(x)\vec{B})/f_{PQ}\pi
+\mbox{``matter''}\,\,\,.
\end{equation}

The last term ``matter'' denotes electric charges of ordinary matters.
The first term in the right hand side 
represents an electric charge made of the axion. 
Although the axion is neutral, it can possess an electric charge 
under a magnetic field owing to the interaction eq(\ref{int}).
The interaction mixes electric and magnetic sectors.
Thus, the axion star is electrically polarized 
under the magnetic field.
( This polarization is oscillating
and so there exist a corresponding oscillating current, 
$J_a=c\,\alpha\,\partial_ta(x)\vec{B}/f_{PQ}\pi$; it can be 
read by checking other Maxwell equations.
Thus, radiations are emitted by the AS, which might be observable. 
We will discuss it in later section. )
Accordingly,  
the electric field $\vec{E_a}$ is induced;  

\begin{equation}
\vec{E_a}=-c\,\alpha \,a(x)\,
\vec{B}/f_{PQ}\pi\,\,\,,
\end{equation}
with $\alpha\simeq 1/137$.
Numerically, its strength is given by 

\begin{equation}
\label{e}
E_a \sim 10^{15}\,\,\mbox{eV}\,\,\mbox{cm}^{-1}\,\,B_{12}\,m_9\, 
\quad ,
\end{equation}
with $B_{12}=B/10^{12}$ G,
where we have used the solution in eq(\ref{a}) for the critical mass. 
Obviously, the spatial extension of the field is given by 
the radius $R_a\simeq 1.6\times 10^{5}\,m_9^{-1}\,\mbox{cm}$ of the AS.
Therefore, we find that the electric field is much strong around a 
neutron star with magnetic field $\sim 10^{12}$ G. 
This electric field is an engine of UHECRs in our model.

It should be mentioned that the existence of the electric field is
originated in the coherence of axions composing axion star. Incoherent
axions can be converted into incoherent photons under a magnetic field. 
But, they never induce
electric field with large spatial extension. This electric field
causes UHECRs and GRBs.
Thus, it is essentially important
in our discussion that the axions are coherent in axion stars.

We should also mention that electromagnetic radiations from the axion
star are caused by the interaction\ref{int} which transforms
axions to photons under 
magnetic field. This is the same phenomenon as one in Sikivie-type
axion detectors\cite{s}; incoherent axions are converted into 
photons in the detector.  
In the case of coherent axions, oscillating electric 
current is induced due to the oscillation of the coherent axions
themselves, as we have shown above. Hence, it is very general property 
for the solutions of axion star to oscillate in time.

\section{generation mechanism of ultra high energy cosmic rays}
The ultra high energy cosmic rays with energy $>10^{20}$ eV pose a 
serious challenge for conventional theories of origin of cosmic rays.
Such high energy cosmic rays can not travel in distance of 
more than $50$ Mpc due to the interaction between the cosmic rays and 
the cosmic background radiation, or infrared background.
But there are no candidate astrophysical sources like 
active galaxy nuclei in the directions of the UHECRs.
Eight events of UHECRS have been detected and the detection rate is 
approximately once per year and per $100\,\mbox{km}^2$.
There are no conventional astrophysical and particle physical explanation
for these observations.

Here we explain the events with a mechanism of
the collision between axion stars and
neutron stars with strong magnetic field $>10^{12}$ Gauss. Under such a
strong magnetic field the axion star induces strong electric field,
which can accelerate charged particles to the energy $>10^{20}$ eV.
The electric field induced in the axion star
is oscillating with the frequency, 
$m_a/2\pi\simeq 2.4\times 10^5\,m_9\,\mbox{Hz}$.
Thus a particle with electric charge Ze can be accelerated by 
the field in a direction 
within the half of the period,
 $\pi/m_a$ or
to a distance $\simeq R_a$ 
( $\sim \pi/m_a\times\mbox{light velocity}$ ), 
unless it collides with other particles within the period.
Thus, the energy $\Delta E$ obtained by the particle is given by

\begin{equation}
\label{delta}
\Delta E=\mbox{Ze}\,E_a\times \pi/m_a
\times \mbox{light velocity}\sim 10^{20}\,\mbox{Z}\,\,\mbox{eV} B_{12}\quad .
\end{equation}
Therefore, the electric field  
can accelerate the charged particle to the energy
$\sim 10^{20}\,\mbox{Z}$ eV. 
These charged particles may be 
baryons contaminated in AS or electron-positron pairs produced 
by the decay of the electric field itself.
Note that the direction of the acceleration is parallel to the
magnetic field. Thus the energy lose of the accelerated particles
due to synchrotron emission is not important 
in the acceleration region.

Comment is in order. It seems apparently from eq(\ref{delta})
that stronger magnetic fields 
yield cosmic rays with higher energies. 
Stronger magnetic fields, however, induce stronger electric fields, which
are unstable against electron-positron pair creations. Therefore, strong
magnetic fields do not necessarily yield cosmic rays with higher energies.

If the particles collide with other particles on the way of acceleration,
in other words, their mean free paths are shorter than $R_a$,
they can not obtain such high energies. It is easy to see, however, that 
the mean free paths of quarks or leptons with much higher energies 
than their masses are longer than 
$R_a$ in magnetosphere of neutron star.
This is because since  
cross sections, $\sigma$, of quarks or leptons with such energies $E$
behaves such as $\sigma \sim 1/E^2$, 
mean free paths $\sim 1/n\sigma$ is longer than
$R_a\sim 10^5$ cm unless number density $n$ 
of particles around the AB is extremely large ( i.e.
$n>10^{44}/\mbox{cm}^3$ for $E=10^{20}$ eV ). 
Obviously, under the electric field eq(\ref{e}),
the particles can obtain energies higher 
than their masses $\sim 1$ MeV by moving parallel to the 
electric field only in the distance 
of $\sim 10^{-9}$ cm.  

The point in the acceleration is that as the electric field is much 
strong, charged particles can gain much energies by moving in 
a short distance. The mean free path becomes larger as  
the particles gain more energies. Thus, under the strong 
electric field the particles can gain effectively the energies without
losing their energies.

As is well known, the strong electric field is unstable\cite{vacdecay} against 
electron-positron pair creations. Thus the field decays when 
the magnetic field of the neutron star is strong sufficiently.
This implies that AS itself decays into the pairs.

Let us estimate the decay rate of the field
and show that the AS decays before colliding directly with a neutron star 
whose magnetic field at surface is stronger than $10^{12}$ G.
We also show that  
the AS can collide directly with a neutron star 
whose magnetic field is relatively weak
$\leq 10^{10}$ G.

The decay rate $R_d$ of the field per unit volume and 
per unit time is given by\cite{vacdecay}

\begin{equation}
\label{Rd}
R_d=\frac{\alpha E_a^2}{\pi^2}\sum_{n=1}^{\infty}
\frac{\exp{(-m_e^2\pi n/eE_a)}}{n^2}
\end{equation}
where   
$m_e$ denotes electron mass. The rate is very small for 
an electric field much weaker than $m_e^2\pi\sim 4\times 10^{16}$ eV/cm.  
The electric field of AS, however, can be very strong 
and it can be comparable to $m_e^2\pi$. Therefore,
the rate is much large. 
Numerically, it is given by 

\begin{equation}
\label{cri}
R_d\sim 10^{47}\,B_{12}^2\,m_9^2\,\, \mbox{cm}^{-3}\mbox{s}^{-1}\,
\sum_{n=1}^{\infty}\exp(-0.7\times10^{2}\,n/B_{12}\,m_9)/n^2\,\,\,.
\end{equation}
Since the spatial extension of the field is approximately given by 
$10^5\,m_9^{-1}$ cm, the total decay rate $W$ of the field in AS is 
$\sim  10^{62}\,B_{12}^2\,m_9^{-1}\mbox{s}^{-1}\,\,
\sum_{n=1}^{\infty}\exp(-0.7\times 10^{2}\,n/B_{12}\,m_9)/n^2$. 
Numerically, it reads

\begin{equation}
\label{W}
W\simeq  10/\mbox{s}\,\,\mbox{for}\,\, B_{12}=0.5,
\quad W\simeq 10^6/\mbox{s}\,\,\mbox{for}\,\,B_{12}=0.55,
\quad W\simeq  10^{32}/\mbox{s}\,\,\mbox{for}\,\,B_{12}=1
\end{equation}
with $m_9=1$.

Therefore, we find that the AS decays very rapidly ( or almost suddenly )
when it approaches a region 
where the strength of the magnetic field reaches a critical value of 
about $10^{12}$ G. Hence, the AS evaporates before colliding with the neutron 
star whose magnetic field at the surface is stronger than
$10^{12}$ G.  
The whole energy of the AS is transmitted to 
electron-positron pairs, each of which can obtain energies $\sim 10^{20}$ eV. 
These particles are emitted into a cone with
very small solid angle. They form 
an extremely short pulse whose width being less
than millisecond. Actually, when we suppose 
that the relative velocity of the AS is equal to light velocity,
it decays approximately within a period of 
$10^{-4}\mbox{sec}\sim \,\, 10^{-5}$sec; 
in the period it passes the region where
the magnetic field increases from 
$0.5\times 10^{12}$ G to $10^{12}$ G.  
These leptons may be converted into baryons and photons through 
the interactions with themselves, interstellar
medium or ejection of progenitor of the neutron star. In particular,
the energies of the leptons are transmitted mainly to 
those of photons. Therefore, high energy photons are 
emitted as ultra high energy cosmic rays in the collision, 
when the magnetic field is sufficiently strong.

On the other hand, high energy protons are emitted
in the collision between neutron star and axion star with
matter contamination. Such a matter can be accelerated to the
ultra high energy before the axion star decays. 
The amount of the protons emitted as UHECRs
depends on the matter contamination in the axion star.
We expect that almost of the same amount of protons 
with that of the leptons are emitted in such a collision.

We comment that the velocity of AS trapped gravitationally to
a neutron star, is 
approximately given by the light velocity 
just when it collides with the neutron star.
This is because an AS is trapped to a neutron star 
when the AS approaches it within a distance $\sim 10^{11}$ cm 
where its potential energy dominates over its kinetic energy. After that, 
the AS goes around the neutron star,
losing its potential energy and angular momentum
by emitting gravitational and electromagnetic waves. 
Finally, the AB collides with it.
Thus, the velocity of the AS
reaches approximately the light velocity 
when it collides with the neutron star.

We have to examine the number density of the leptons produced during the decay
of AS and have to check whether or not 
the density is less than $10^{45}/\mbox{cm}^3$;
the particles can not obtain the ultra high energies $\sim 10^{20}$ eV 
unless the number density surrounding AS
is not beyond the value quoted. They collide
with other particles on the way of the acceleration and lose their energies.
Suppose that the mass $M_a$ of AS is transmitted to 
the energy of electron-positron pairs. Then,  
their number is given by $\sim M_a/m_e$.
They may be produced  
in the volume $R_a^3$ of AS. 
If we assume that they remain to stay in the volume after their production,
the number density is $\sim M_a/m_e\,R_a^3\simeq 10^{44}/\mbox{cm}^3$. 
Actually, the pairs escape in a direction pointed by the electric field
within the life time of the field $10^{-4}\,\mbox{sec}\sim 10^{-5}$ sec.
Thus, the real number density in AS is less than $10^{44}/\mbox{cm}^3$.
Therefore, we find that the number density of the pairs produced from
the decay of AS is not so large to interrupt them obtaining the ultra high 
energies.

We can see from eqs(\ref{delta}) and (\ref{cri}) 
that the critical electric field depends on the factor of $B_{12}\times m_9$,
while the energy $\Delta E$ obtained by 
accelerated charged particles depends only on the factor of $B_{12}$.
Therefore, we find that if axion mass $m_a$ is smaller than 
$10^{-9}$ eV, the maximal energy of cosmic rays can be 
larger than $10^{20}$ eV when a neutron star has strong magnetic field 
$B>10^{12}$ G. For instance, if $m_a=0.2\times 10^{-9}$ eV,
the cosmic rays with energies $\sim 5\times 10^{20}$ ZeV can be produced
when a neutron star has magnetic field $>5\times 10^{12}$ G at the surface.

\section{generation mechanism of gamma ray bursts}
Gamma ray bursts are observed daily from sources extending out
to those of the most distant galaxies in the Universe. Duration of 
the bursts ranges from millisecond to thousand seconds. Pulse shape
structures vary much from bursts to bursts. 
Typical energies of the gamma rays observed are of the order of $0.1$ MeV.
A mysterious puzzle of these bursts 
is the huge amount of energies released in the bursts.
Some of the bursts carry energies $\sim 10^{54}$ erg when spherical 
explosion from sources is assumed. Furthermore, baryon contamination
in fire balls of the explosion is less than $10^{-4}M_{\odot}$. 
Therefore, it is very difficult to make realistic astrophysical models 
for explaining these properties, although it is believed optimistically 
that the origin of these phenomena would be some known astrophysical
objects, such as merger of neutron star - neutron star, collapse of
extremely massive star, etc.

Here, we explain a particle physical 
generation mechanism of gamma ray bursts. 
According to the mechanism, we can understand 
the problems of the huge energies released and the  
baryon contamination. But it is still difficult to understand the variant
of the durations and the pulse shapes.
The bursts are possibly produced in the collision between AS and neutron star
with relatively small magnetic field such as $\leq 10^{10}$ G.
In such a weak magnetic field,
the AS collides directly with the neutron star and dissipates its whole energy
in an outer crust of the neutron star. Actually,
the decay rate $W$ of the electric field is negligibly small for the case of 
the weak magnetic field $\leq 10^{10}$ G. 
Thus the AS does not decay before colliding directly with 
such a neutron star. The AS, however, decays in magnetized conducting medium
such as the outer crust of neutron star. Namely the electric field of the AS
induces electric current in the crust, which dissipates its energy owing to
finite electric conductivity of the crust. Thus, the AS decays very rapidly 
by dissipating its energy. Consequently, 
particles of the crust are emitted as a jet
in the collision, which forms fireball emitting gamma ray bursts.

We will estimate the rate of the energy dissipation
in the outer crust with use of the conductivity, $\sigma=10^{26}$/s\cite{con}.
The value of the conductivity, in general, depends on physical parameters of the crust 
such as temperature, composition, density, etc.. The value of $\sigma$
we use, however, is a typical one of the crust and does not vary so much even if 
we change the parameters of the density, temperature, or composition
within a reasonable range.

We calculate\cite{iwazaki,iwa} the rate $W_{dis}$ of 
the dissipation per unit time 
and unit volume as follows,

\begin{eqnarray}  
W_{dis}&=&\int_{AS}{\sigma E_a^2}d^3x/\int_{AS}=
4c^2\times 10^{61}/(4\pi/3R_a^3)\,\,\mbox{erg/s}\,\frac{\sigma}{10^{26}/s}\,
\frac{M_a}{10^{-1}M_{\odot}}\,\frac{B^2}{(10^{10}\mbox{G})^2} \\
&\simeq& 2.4\times 10^{45}\mbox{erg/s}\,\mbox{cm}^3\,
\frac{\sigma}{10^{26}/s}\,
\frac{B^2}{(10^{10}\mbox{G})^2}\,m_9^2\,,
\end{eqnarray}
where we have used the solution in eq(\ref{a}) for the critical mass. 
The integration has been performed over the volume ( $=4\pi R_a^3/3$ ) 
of an AS, which has been supposed to be 
involved completely in the conducting medium.
Thus it represents an average rate of the energy dissipation in the outer crust.

We find that the energy dissipation
proceeds very rapidly in the medium.
We compare the rate with the energy density of the AS, which is
given by $10^{37}\mbox{erg}/\mbox{cm}^3$. Hence, we find that 
the AS dissipates 
the whole energy $\sim 10^{53}$ erg in the outer crust 
even if it rushes into the medium with the light velocity;
the depth of the crust is equal to several hundred meters.
It never reaches the core of the neutron star.

This very rapid energy release leads to the ejection of the particles composing
the crust.
The ejection could be emitted
into a cone with small solid angle as a jet. This is because the particles 
( mainly irons ) of the neutron star are accelerated and emitted
in the direction parallel to 
the strong electric field $\sim 10^{13}\,B_{10}$ eV/cm. 
The particles are accelerated to energies $\sim 10^{18}$ eV, while 
their characteristic energies inside the crust are $\sim 10^6$ eV.
Thus, we expect that solid angle of the jet is much small,
although the particles do not necessarily obtain such high energies
because of loosing energies by the collisions with others.  
The fact that the whole energy of AB is dissipated only in the 
outer crust, implies that the ejection are only particles composing the 
crust. Thus a fraction of the baryon contamination in the jet is less than
$10^{-5}M_{\odot}$ as required observationally. We expect 
that a large amount of neutrinos with high energies 
is also produced in the collision with the rapid 
energy dissipation. The neutrinos burst can be observed
in future observatory although it has lost the direction pointing to 
a source of a GRB in which it was born.
This is our generation 
mechanism of gamma ray bursts with energies $\sim 10^{53}$ erg.

In the above case, the AS dissipates its whole energy  
in the first collision. On the other hand, an AS may collide several times
with a neutron star when its mass is much smaller than the critical mass
$\sim 10^{-1}M_{\odot}$; the mass has been assumed 
as a characteristic mass scale of ABs.
We see from the general formula eq(\ref{R}) of $R_a$ that the radius of the AS 
with smaller mass than 
the critical one is larger than the critical radius $\sim 10^5$ cm. 
For example, if its mass is given by $10^{-2}M_{\odot}$, the radius is about
$10^{6}$ cm. This is comparable to the radius of the neutron star.
Hence, the collisions may occur several times.
Such an AS never loose its whole energy in the first collision.
How many times the collisions occur depends on the 
collision parameters. Unless the centers of both stars are on the straight 
line parallel to the relative velocity, the whole energy of AS does not 
dissipate in the first collision. There could be several collisions 
after the first collisions. 
The electric field induced in the AS is also small compared 
with the one mentioned above; 
$E_a\sim 10^{11}\,\,\mbox{eV}\,\,\mbox{cm}^{-1}\,\,B_{10}\,m_9$,
for $M_a=10^{-2}M_{\odot}$. Thus, the energies 
of the particles emitted from the neutron star are much smaller than $10^{18}$
eV. Furthermore, since the rate of the energy dissipation becomes smaller in 
the case of AS with smaller mass, the rate of the energy release is
also smaller. It means that jet emitted in the collision
is softer; its flux ( and energies of each particles ) is smaller 
than that of 
a jet emitted from the collision of AS with the critical mass. 
The total energies released ( $\sim 10^{52}$ erg for $M_a=10^{-2}M_{\odot}$ ) 
are also smaller than $10^{53}$ erg.
We expect that these collisions generate
gamma ray bursts with long duration and soft gamma rays. On the other hand,
gamma ray bursts with short duration and hard gamma rays are produced in  
the first collision in which the whole energy is dissipated completely.
 
We should comment that when the axion star with much matter contamination 
collides, the jet ejected from the collision 
may contain baryon contamination much larger than $10^{-5}M_{\odot}$.
Thus in such a collision, Lorentz factor of the jet 
is not so large that gamma ray bursts may not be produced.

\section{rate of energy release in uhecrs}
We now wish to estimate an energy release rate in the collisions between
ASs and neutron stars. We assume that UHECRs are produced in the 
collision of neutron stars with relatively
strong magnetic field $>10^{12}$ G and that GRBs are produced in the
collision of neutron stars 
with relatively weak magnetic field $\leq 10^{10}$ G.
Thus it is assumed that all of neutron stars are divided
into those with $B>10^{12}$ G and those
with $B\leq 10^{10}$ G. We take the fraction of the neutron stars 
with $B>10^{12}$ G to be $f$; $0<f<1$. 
Although the collisions may occur, in general, in the whole
Universe, we estimate the rate of the collision which
arises in a galaxy. Thus the real rate is larger than a rate in
our estimation.
Furthermore,
we assume that the dark matter is composed mainly of 
axion stars. In particular a hale is composed of 
the axion stars. 
The estimation, however, involves several ambiguities associated with
number density of neutron stars, energy density of dark matter 
or velocity of ASs
in the Universe etc. Therefore the estimation does not lead to 
a conclusive result although our result is consistent with
the observations\cite{uhe}.

First we assume that almost of all neutron stars are 
mainly located around a galaxy 
in which they were born.
We also assume that their distribution
is uniform in a halo of the galaxy whose size is supposed to be 
given by $50$kpc. Our result does not depend
on the precise value of the size although it depends on 
mean energy density of the dark matter in the halo. 
Thus the collision between neutron star and 
axion star arises in the halo. The number of the neutron stars
located around a galaxy is supposes 
to be $\sim 10^{9}$; the present rate of appearance 
of supernovae in a galaxy is about $0.1\sim 1$ per $10$ year and 
the rate of the appearance could be 
larger in early stage of the galaxy than the one at present.
Therefore, the number density $\rho$ of the neutron stars
would be given such that $\rho \simeq 10^9/(50\mbox{kpc})^3$.

v

All of these neutron stars are assumed to possess either  
strong magnetic field $>10^{12}$ G or weak magnetic field $\leq 10^{10}$.
Furthermore,
to estimate the rate of the energy release 
we need to know average density $\epsilon$ 
of the dark matter, i.e. density of axion stars.
Here we use a value\cite{text} of 
$\epsilon\simeq 0.5\times 10^{-25}\mbox{g}/\mbox{cm}^3$,
which represents a local density of our halo:
Our result depends on $\epsilon$ linearly. 
Using these parameters we can estimate the rate owing to 
the collision between the AS 
and the neutron star. The collision takes place as a result of 
AS losing angular momentum and potential energy soon
after AS being trapped
gravitationally to a neutron star. Thus, we estimate
the cross section for 
a neutron star to trap an AS 
in the following. Namely an AS is trapped by the neutron star when
the AS approaches the neutrons star within a distance $L_c$ in which
its kinetic energy $M_av^2/2$ is equal to its potential energy 
$\,1.5\times M_{\odot}\,M_a\,G/L_c$ around the neutron star.
$G$ is gravitational constant. Here 
the mass of the neutron star and the relative velocity $v$ are assumed to be 
$1.5\times M_{\odot}$ and $3\times 10^7$ cm/s, respectively. 
Thus, the cross section is found such as $L_c^2\pi$ 
with $L_c\simeq 6\times 10^{11}\,\mbox{cm}$.
The lose of its potential energy and angular momentum would be caused by 
the emission of gravitational and electromagnetic waves;
electromagnetic radiations arise due to the oscillating current $J_a$.
It also takes place by the interaction 
between the electric field
of AS and plasma wind from neutron star. The detail 
calculation has not yet been done.
It will be clarified in future publication.

Anyway, under the assumptions,
we obtain the rate of the collision between 
neutron star and axion star per Mpc$^{3}$ and per year,

\begin{equation}
(\epsilon /10^{-1}M_{\odot}\times \rho \times L_c^2\pi\times v\times 1\, 
\mbox{year})/(20)^3\simeq 3\times 10^{-10}\,/\mbox{Mpc}^3\,
\, \mbox{year}.
\end{equation}
where we have obtained the rate per a galaxy with its volume $(50\mbox{kpc})^3$ 
and have divided it with $(20)^3$ to get the rate per Mpc$^3$.
This rate is divided into the rate of production of UHECRs and 
the rate of production of GRBs.

Here we comment that the rate heavily depends on 
the relative velocity $v$: $L_c^2\times v \sim v^{-3}$. Thus if we
choose a value $v=2 \times 10^{7}$ cm/s, the rate is given by
$\sim 10^{-9}\,/\mbox{Mpc}^3\,\,\mbox{year}$. Thus our result involves
an ambiguity of the order of $10$. 

Since the energy of $\sim 10^{53}$ erg is released in the collision, 
we find that the rate of the energy release as UHECRs is given by 

\begin{equation}
\sim ( 3\times 10^{43}\sim 10^{44})\,f\, \mbox{erg}/\mbox{Mpc}^3\,\,\mbox{year}.
\end{equation} 
where we assumed that the whole energy released is 
transmitted to UHECRs or GRBs. Our result 
does not include the contributions from 
collisions arising far outside of a galaxy. The real rate is 
larger than the rate in our estimation. Thus we may conclude that
the result roughly agrees with the observation.  It is well known that
both rates of energy release in UHECRs and GRBs are nearly the same
with each other.
It implies that the fraction of the neutron stars with $B>10^{12}$ G
is nearly the same with the fraction of neutron stars with $B\leq 10^{10}$.
Taking account of several ambiguities in the 
parameters used above and the rough estimation, we can say simply that
our model can explains roughly 
the observations\cite{uhe}. 

\section{observational signal of axion stars}
We have discussed our model for generation mechanisms of UHECRs and GRBs.
These are generated in the collision between axion star and neutron star.
An significant question is how we can check observationally our model.
Since we have supposed that dark matter in Universe is composed mainly 
of axion stars, observation of the axion stars is important for 
checking validity of our model. There are several
attempts of the detection of the axion itself. These attempts use axions 
conversion into photons under magnetic field in terrestrial experiments. 
But there are no attempts to observe
axion stars. Since a collision between an axion star and
the earth is very rare, we need to detect signals from 
the axion stars located far away from the earth.

Here we wish to discuss the observation of two types of signals 
from the axion stars.
One is associated with gravitational lensing and the other one associated
with monochromatic radiations from the ASs. 

It was very intriguing to detect microgravitational lens effects
associated with the halo of our galaxy. Since the first observations
have been reported, 
several candidates for MACHOs have been proposed. Observationally, 
it is necessary for the candidates to have masses $M$ roughly such as 
$0.1\times M_{\odot}< M < 0.7\times M_{\odot}$. 
Since we have assumed that the halo of our galaxy is composed of ASs and
their mass is $\sim 10^{-1}M_{\odot}$, 
the ASs are plausible candidates 
for the MACHOs.
As has been recognized, 
baryonic candidates like white dwarfs, neutron stars, etc. have serious 
problems. Nonbaryonic candidates are favored\cite{non}. 
The problems are associated with 
chemical abundance of carbon and nitrogen in the Universe: If these baryonic 
stars are MACHOs, 
an overabundance of the elements is produced far in excess of what
is observed in our galaxy.
Hence, the ASs are theoretically the most fascinating candidates for
the MACHOs as nonbaryonic ones. 
They are also candidates for the generators of 
both UHECRs and GRBs.
If the most favorable mass of the MACHO is  
$0.5M_{\odot}$,
we need to choose 
$m_a\simeq 0.2\times 10^{-9}$ eV 
since the mass of the ASs is given by $\simeq 10^{-1}M_{\odot}/m_9$. 
We note that smaller axion mass leads to stronger electric field as 
well as larger mass of AS. Thus,
it yields 
higher energies of the UHECRs, $\sim 5\times 10^{20}$ eV 
and of GBRs, $\sim 5\times 10^{53}$ erg than the ones 
we have claimed above. Accordingly, 
the determination of the mass of MACHOs 
gives the upper limit of both the energies of the ultra high energy 
cosmic rays and the energies released in the gamma ray bursts in our
mechanism.

We point out another way of the observation of the axion stars. Since 
the electric field $E_a$ as well as electric current $J_a$ induced in ASs is 
oscillating, 
electromagnetic radiations
are emitted\cite{iwasol,rad}; 
the frequency of these monochromatic radiations is 
$\simeq 2.4\times 10^5\,m_9$ Hz ( their wave length is
given by $\sim 1.3\times 10^5/m_9$ cm. ) 
Their flux is stronger as the magnetic field imposed on AS is stronger.
AS revolves around neutron star, loosing its potential
energy and angular momentum, and then collides with it. 
In the revolution, AS emits the radiations with flux depending
on the strength of the magnetic field. 
Hence, we expect that such radiations with maximal flux can be detected 
just in advance of UHECRs. Their flux is strongest among others just before 
UHECRs being emitted; the magnetic field is strongest when 
the AS decays into charged particles. 
It is easy to 
estimate their luminosity\cite{rad} by evaluating 
gauge potential $A_i$,

\begin{eqnarray}
A_i&=&\frac{1}{R_0}\int_{AS} J_a(t-R_0+\vec{x}\cdot\vec{n})\,d^3x
=\frac{c\alpha\sigma a_0B_i}{\pi R_0}\int_{AS}
\sin m_a(t-R_0+\vec{x}\cdot\vec{n})\,\exp(-r/R_a)d^3x\\
&=&\frac{4c\alpha m_a a_0B_i}{R_0}\,\frac{R_a^3}{(m_a^2R_a^2+1)^2}\,
\sin m_a(t-R_0)\,,
\end{eqnarray}
where $R_0$ ( $\vec{n}$ ) denotes a distance ( direction )
between ( from ) the neutron star and ( to ) the earth.    
We have integrated the current $J_a$  over the AB, which is located 
near the neutron star. Thus, 
the luminosity $L$ of the radiations  
is given by

\begin{eqnarray}
L&=&4 c^2 \alpha^2 a_0^2 m_a^4 B^2 \frac{R_a^6}{(m_a^2R_a^2+1)^4}\\
&\simeq& 6.7\times 10^{41}\,B_{12}^2\,\mbox{erg/s},
\end{eqnarray}
with $c=1$.
If the emission arises 
in a distance $\sim 10$ Mpc from the earth, we obtain the flux at the earth,
$\sim 10^9\,\mbox{Jy}\,B_{12}^2/m_9$; we have assumed that 
the velocity of the AS revolving 
is $\sim 0.1 \times $ light velocity. 
Although the possibility of observing the radiations 
is very intriguing,
it might be difficult to
detect the radiations with such a low 
frequency because they could be absorbed by interstellar 
ionized gases before arriving the earth. 
v

\vskip .5cm

The author wishes to express his thank to all the staffs in Theory Group,
Tanashi Branch, High Energy Accelerator Research Organization 
for their warm hospitality extended to him.
This work is supported by the Grant-in-Aid for Scientific Research
from the Ministry of Education, Science and Culture of Japan No.10640284


\end{document}